\documentclass[12pt]{article}
\usepackage{mathrsfs}
\usepackage{amssymb}
\usepackage{dsfont}
\normalsize
\input{epsf}

\begin{document}
\addtolength{\baselineskip}{.20mm}
\newlength{\extraspace}
\setlength{\extraspace}{2mm}
\newlength{\extraspaces}
\setlength{\extraspaces}{2mm}

\newcommand{\newsection}[1]{
\vspace{15mm} \pagebreak[3] \addtocounter{section}{1}
\setcounter{subsection}{0} \setcounter{footnote}{0}
\noindent {\Large\bf \thesection. #1} \nopagebreak
\medskip
\nopagebreak}
\newcommand{\newsubsection}[1]{
\vspace{1cm} \pagebreak[3] \addtocounter{subsection}{1}
\addcontentsline{toc}{subsection}{\protect
\numberline{\arabic{section}.\arabic{subsection}}{#1}}
\noindent{\large\bf 
\thesubsection. #1} \nopagebreak \vspace{3mm} \nopagebreak}
\newcommand{\ba}{\begin{eqnarray}
\addtolength{\abovedisplayskip}{\extraspaces}
\addtolength{\belowdisplayskip}{\extraspaces}

\addtolength{\belowdisplayshortskip}{\extraspace}}

\newcommand{\be}{\begin{equation}
\addtolength{\abovedisplayskip}{\extraspaces}
\addtolength{\belowdisplayskip}{\extraspaces}
\addtolength{\abovedisplayshortskip}{\extraspace}
\addtolength{\belowdisplayshortskip}{\extraspace}}
\newcommand{\ee}{\end{equation}}
\newcommand{\STr}{{\rm STr}}
\newcommand{\figuur}[3]{
\begin{figure}[t]\begin{center}
\leavevmode\hbox{\epsfxsize=#2 \epsffile{#1.eps}}\\[3mm]
\parbox{15.5cm}{\small
\it #3}
\end{center}
\end{figure}}
\newcommand{\im}{{\rm Im}}
\newcommand{\calm}{{\cal M}}
\newcommand{\call}{{\cal L}}
\newcommand{\sect}[1]{\section{#1}}
\newcommand\hi{{\rm i}}
\def\bea{\begin{eqnarray}}
\def\eea{\end{eqnarray}}

\begin{titlepage}
\begin{center}

\vspace{3.5cm}

{\Large \bf{Noether Symmetry Approach in ``Cosmic Triad" Vector Field Scenario}}\\[1.5cm]

{Yi Zhang $^{a,b,}$\footnote{Email: zhangyia@cqupt.edu.cn},}{Yun-Gui
Gong $^{a,}$\footnote{Email: gongyg@cqupt.edu.cn}, }{Zong-Hong Zhu
$^{b,}$\footnote{Email: zhuzh@bnu.edu.cn},} \vspace*{0.5cm}

{\it $^{a}$College Mathematics and Physics, Chongqing Universe of Posts and Telecommunications, \\ Chongqing 400065, China

$^{b}$ Department of Astronomy, Beijing Normal university,
\\  Beijing 100875, China}

\date{\today}
\vspace{3.5cm}

\textbf{Abstract} \vspace{5mm}

\end{center}
To realize the accelerations in the early and late periods of our
universe, we need to specify potentials for the dominant fields. In
this paper, by using the Noether symmetry approach, we try to find
suitable potentials in  the ``cosmic triad" vector field scenario.
Because the equation of state parameter of dark energy has been
constrained in the range of $-1.21\leq \omega\leq -0.89$ by
observations, we derive the Noether conditions for the vector field
in quintessence, phantom and quintom models, respectively. In the
first two cases, constant potential solutions have been obtained.
What is more, a fast decaying point-like solution with power-law
potential is also found for the vector field in quintessence model.
For the quintom case, we find an interesting constraint
$\tilde{C}V_{p}'=-CV_{q}'$ on the field potentials, where $C$ and
$\tilde{C}$ are constants related to the Noether symmetry.

\end{titlepage}

\section{Introduction}\label{sec1}
The inflationary paradigm was used as a resolution of the problems, such as horizon and flatness problems
in standard cosmology \cite{Guth:1980zm}. The dark
energy scenario was proposed to explain the current accelerating expansion of the universe found by
the type Ia supernova observations \cite{Riess:1998cb}.
Various candidates are suggested to explain the early and  current
accelerations. Assuming that a scalar field provides the driving force of the acceleration
is the most popular and economic explanation.
However, vector field scenarios are workable as well, which also have the advantage that
their basic particle is present in nature. The vector field inflationary
scenario was firstly proposed by Ford with the characteristic of a natural large scale anisotropy
\cite{Ford:1988wq}. Nevertheless, now, at least  three methods
\cite{Zhang:2009tj} could solve the anisotropic problem: they are
the time-like vector field scenario \cite{Kiselev:2004py},
the``cosmic triad" vector field scenario \cite{Bento:1992wy} and the
``N-flation" vector field model \cite{Golovnev:2008cf}.
As
important applications  in the early universe, the vector field  can be used in inflation \cite{Kanno:2006ty},  used as
a curvaton \cite{Dimopoulos:2005ac}, the stability is also discussed \cite{Himmetoglu:2008zp}. What is more,
vector field is one of the p-form fields \cite{Germani:2009iq}
and can be identified as  the electricmagnetic field \cite{Hosotani:1984wj}.
The vector field scenario  is a worthy topic to work on.

However, these paradigms could  not be  satisfactorily established
without considering their connection with a fundamental theory.
Therefore, we have to face the problem of choosing suitable
potentials from fundamental physics.
 The Noether symmetry approach
has been revealed as a useful tool to find out exact solutions in
cosmology. It is also an effective method to select models motivated
at a more fundamental level.
 By choosing the constant of motion, Noether symmetry  reduces the
dynamical system. In most  cases, results are integrable because of
the conserved quantities. Results of  previous works,
which  concern the scalar
 field cosmology \cite{deRitis:1990ba,zhang,various model} and the $f(R)$ cosmology \cite{f(R)},
 are encouraging.
Then, it is natural to ask what potential will be obtained in vector
field scenario by applying the Noether symmetry approach. In the
present paper, we try to give an answer. First of all, among the
above three vector field scenarios, we choose the ``cosmic triad"
model to discuss,  which  is easy to deal with.

Observations suggest that the dark energy equation of state (EoS)
parameter is in the range of $-1.21\leq \omega\leq -0.89$
\cite{Riess:2004nr} which  has a possibility of crossing the phantom
divider $\omega=-1$. Although phantom type of matter with negative
kinetic energy  has well-known problems, it was implicitly suggested
by astronomical observations and
 has also been widely
studied as  dark energy. It is phenomenologically significant
 and worthy of putting other theoretical difficulties aside temporally. Therefore,
 for a complete description,  corresponding to the classification of the scalar fields,
 we discuss three types of matter in ``cosmic triad"  vector field scenario,
 which are the quintessence type of vector fields with positive kinetic terms,
 the phantom type of vector fields with
 negative kinetic terms, and the quintom type of vector fields with both
 positive and  negative kinetic terms
 \footnote{ The scalar field quintessence with positive kinetic term is proposed by Ref.\cite{Ratra:1987rm}; the scalar field phantom with
   negative kinetic term is suggested by Ref.\cite{Caldwell:1999ew}; and
 the quintom  with both positive and  negative kinetic terms is proposed by Ref.\cite{Feng:2004ad}. And with the help of modified gravity,
  a lot of models could cross $\omega=-1$ as well, for example in \cite{cross-1}.}.

 The paper is organized as follows. In section \ref{sec2}, we introduce the ``cosmic triad" vector field model.
 In section  \ref{sec3}, we introduce the  Noether symmetry  approach, and apply it to both quintessence and phantom cases.
 We perform the change of  variables and get the solutions.
In section  \ref{sec4}, we discuss the application of Noether
symmetry approach  in  vector field quintom  case. In section \ref{sec5},  we draw our conclusions.

\section{``Cosmic Triad" Scenario}\label{sec2}
Based on observations, we assume that the geometry of space-time is described by the
flat Friedmann-Robertson-Walker (FRW) metric,
\begin{eqnarray}
 ds^{2}=-dt^{2}+a^{2}(t)\sum^{3}_{i=1}(dx^{i})^{2},
\end{eqnarray}
where $a$ is the scale factor.  Meanwhile, as we consider both the
vector fields and the dust matter in the system, the total action
is
 \be
 S_{tot}=S_{A}+S_{m},
 \ee
 where $S_{A}$ is the action for vector fields, $S_{m}$ is the action for dust matter. The density of dust  matter can be expressed as $\rho_{m}=\rho_{m0}(a_{0}/a)^{3\gamma}$, where $\rho_{m0}$ is an initial constant and $0<\gamma\leq2$.
Here, we limit our analysis to $\gamma=1$ which corresponds to the
pressureless dust matter with $P_{m}=0$.

The ``cosmic triad"  model \cite{Bento:1992wy}, as a
realistic vector field scenario, which can be derived from  a
gauge theory with $SU(2)$ or $SO(3)$ gauge group, is  proved to be
compatible with the background metric, with the following action
\begin{eqnarray}
   \label{action1}
 S_{A1}=\int d^{4}x \sqrt{-g}\left[\frac{R}{16 \pi G}+\sum^{3}_{a=1}\left(\epsilon\frac{1}{4}F^{a}_{\mu\nu}F^{a\mu\nu}-V(A^{a2})\right)\right],
\end{eqnarray}
where Latin indices label the gauge fields ($a,b=1,2...$), and Greek
indices label the different space-time components
($\mu,\nu=1,2...$). $F^{a}_{\mu
\nu}=\partial_{\mu}A_{\nu}^{a}-\partial_{\nu}A_{\mu}^{a}$ is the
field strength,  $A_{\mu}^{a}$ is the vector filed and
$A^{a2}=g^{\mu\nu}A_{\mu}^{a}A_{\nu}^{a}$. For the purpose of
describing both the positive and negative kinetic terms, we put the
parameter $\epsilon$ in the action. When $\epsilon=1$, the kinetic
term of the ``cosmic triad" is positive, and one has the vector
field quintessence case. When $\epsilon=-1$, the kinetic term of the
``cosmic triad" is negative,  and then one has the vector field
phantom case. The action of the vector field quintom  is
\begin{eqnarray}
   \label{action2}
 S_{A2}=\int d^{4}x \sqrt{-g}\left[\frac{R}{16 \pi G}+\sum^{3}_{a=1}\left(\frac{1}{4}F^{a}_{q\mu\nu}F_{q}^{a\mu\nu}-V_{q}(A_{q}^{a2})\right)+
 \sum^{3}_{b=1}\left(-\frac{1}{4}F^{b}_{p\mu\nu}F_{p}^{b\mu\nu}-V_{p}(A_{p}^{b2})\right)\right],
\end{eqnarray}
where the subscript $q$ denotes quintessence and subscript $p$ denotes  phantom.

Following the assumption in Ref.\cite{Bento:1992wy}, the ``cosmic
triad" scenario becomes compatible  with spatial isotropy by using
the assumptions below for the vector fields.
 For the quintessence and phantom cases,
\be
\label{A1}
A^{b}_{\mu}=\delta^{b}_{\mu}A(t)\cdot a,
 \ee
and for   action (\ref{action2}) which notifies the quintom, it is
 \be
\label{A2} A^{b}_{p\mu}=\delta^{b}_{\mu}A_{p}(t)\cdot
a,\,\,\,A^{b}_{q\mu}=\delta^{b}_{\mu}A_{q}(t)\cdot a,
 \ee
 where $A$, $A_{p}$, $A_{q}$ are scalars.
 The three kinds of vector fields with the same kinetic terms point  in three mutually orthogonal spatial directions and  share the same time-dependent length.
It turns out that this scenario is able to drive a stage of
accelerating expansion in the universe, and exhibits tracking
attractors that render cosmic evolution insensitive to initial
conditions \cite{Bento:1992wy}.


\section{The Noether Symmetry Approach in Quintessence and Phantom Cases}\label{sec3}

As discussed in the introduction, we are looking for solutions
induced by symmetries. Noether symmetry approach is such a powerful
tool that it can find the solution for a given Lagrangian. This
method looks for the related cyclic variables  and
consequently reduces the dynamics of the system to a manageable one.
In  vector field quintessence and phantom cases, we treat  action
(\ref{action1})  as a dynamical system in which the scale factor $a$
and the scalar field $A$ play the role of independent dynamical
variables. Then a configuration space $\mathcal{Q}=(a,A)$ may be
considered. To study the related symmetries, we need an effective
point-like Lagrangian for the model whose variation with respect to
the dynamical variables yields the correct equations of motion.
Based on action (\ref{action1}), the point-like Lagrangian takes
such a form
\begin{eqnarray}
\label{L} \mathcal{L}_{1}=\mathcal{L}_{A1}+\mathcal{L}_{m}=
3a\dot{a}^{2}-\frac{3}{m_{pl}^{2}}\left(\epsilon\frac{a^{3}\dot{A}^{2}+
a A^{2}\dot{a}^{2}+2
a^{2}A\dot{a}\dot{A}}{2}-a^{3}V(A^{2})\right)+\frac{\rho_{m0}}{m_{pl}^{2}},
\end{eqnarray}
where   $m_{pl}^{2}=(8\pi G)^{-1}$ is the Planck mass.

The ``energy function" associated with $\mathcal{L}_{A1}$ is
\begin{eqnarray}
\label{E}
E_{\mathcal{L}_{1}}=\frac{\partial\mathcal{L}_{1}}{\partial
\dot{q_{i}}}\dot{q_{i}}-\mathcal{L}_{1}=3a^{3}\left(\frac{\epsilon(\dot{A}+HA)^{2}}{2}+V(A^{2})+\frac{\rho_{m0}a^{-3}}{3}-m_{pl}^{2}H^{2}\right),
\end{eqnarray}
where $q_{i}$ is the variable $a$ or $A$ in the configuration space.
If considering the vanishing of the ``energy function" as a
constraint, we get the  Friedmann equation
\begin{eqnarray}
\label{E}
H^{2}=\frac{1}{m_{pl}^{2}}\left(\frac{\epsilon(\dot{A}+HA)^{2}}{2}+V(A^{2})+\frac{\rho_{m0}a^{-3}}{3}\right).
\end{eqnarray}
And,  the Euler-Lagrange equations associated with $\mathcal{L}_{1}$
are
\begin{eqnarray}
\frac{d}{dt}(\frac{\partial\mathcal{L}_{1}}{\partial
\dot{q_{i}}})-\frac{\partial\mathcal{L}_{1}}{\partial q_{i}}=0.
\end{eqnarray}
When  $q_{i}=a$, the explicit form of the equation  can be written
down as
\begin{eqnarray}
 \label{me0}
 \dot{H}=-\frac{[2\epsilon(\dot{A}+HA)^{2}+V'A]}{2m_{pl}^{2}},
\end{eqnarray}
which is  called the Raychaudhuri equation, where  $V'=d V/d A$. When
$q_{i}=A$, the Euler-Lagrange equation is
\begin{eqnarray}
  \label{me}
  \ddot{A}+3H\dot{A}+(2H^{2}+\dot{H})A+\epsilon V'=0.
\end{eqnarray}

In the present paper, as both the positive and negative kinetic terms are discussed, it is natural and
  necessary to concern the EoS parameter  which is related to the sign of the kinetic terms tightly.
From the action (\ref{action1}), we can give out the energy
density and the pressure of
 the vector fileds
\begin{eqnarray}
&& \rho_{A}=\frac{3\epsilon}{2}(\dot{A}+HA)^{2}+3V(A^{2}),\\
 && P_{A}=\frac{\epsilon}{2}(\dot{A}+HA)^{2}-3V(A^{2})+V'A,
 \end{eqnarray}\
then we can write down the EoS parameter
 \begin{eqnarray}
\label{w}
\omega_{A}=\frac{P_{A}}{\rho_{A}}=\frac{\epsilon(\dot{A}+HA)^{2}/2-3V(A^{2})+V'A}{3\epsilon(\dot{A}+HA)^{2}/2+3V(A^{2})}.
 \end{eqnarray}
When $\epsilon=1$,  $\omega_{A}<-1$ requires
$2(\dot{A}+HA)^{2}+V'A<0$, and the potential must be ``tachyonic"
which means a negative $V'$. When $\epsilon=-1$, $\omega_{A}<-1$
requires that $-2(\dot{A}+HA)^{2}+V'A<0$, we don't need to make $V'$
smaller than $0$. However, the vector field scenario is different
from the scalar field  case, which  just needs $\epsilon
\dot{\phi}^{2}<0$ if asking $\omega_{\phi}<-1$. Furthermore, the
vector field quintessence and phantom allow that one crosses
$\omega_{A}=-1$.

The Noether symmetry approach consists in considering the two
equations (\ref{me0}) and  (\ref{me}) as a second-order dynamical
system with the following vector field which is an infinitesimal
generator of a point transformation on the  configuration space
$\mathcal{Q}=(a,A)$
\begin{eqnarray}
\label{X}
X=\alpha\frac{\partial}{\partial a}+\beta\frac{\partial}{\partial A}+\dot{\alpha}\frac{\partial}{\partial \dot{a}}+\dot{\beta}\frac{\partial}{\partial \dot{A}},
\end{eqnarray}
 where $\alpha$ and $\beta$ are generic functions of $a$ and $A$. And the  tangent space for the related bundle is $T\mathcal{Q}=(a,A,\dot{a},\dot{A})$. $X$ can be treated as
 a vector field on $T\mathds{R}^{2}$, which is the tangent bundle of $\mathds{R}^{2}$,
 with natural coordinates $(a,A,\dot{a},\dot{A})$.
The Lagrangian is invariant under the transformation $X$ if
\begin{eqnarray}
\label{L1}
L_{X}\mathcal{L}_{1}=\alpha\frac{\partial\mathcal{L}_{1}}{\partial
a}+\frac{d \alpha}{dt}\frac{\partial \mathcal{L}_{1}}{\partial
\dot{a}}+\beta \frac{\partial\mathcal{L}_{1}}{\partial A}+\frac{d
\beta}{dt}\frac{\partial \mathcal{L}_{1}}{\partial \dot{A}}=0,
\end{eqnarray}
where $L_{X}$ stands for Lie derivative with respect to $X$.  We
claim that the dynamical system has Noether symmetries when
$L_{X}\mathcal{L}_{1}=0$. In this way,
 the transformation on the base space  can preserve the second-order character of the dynamical field,
with an explicit expression as
\begin{eqnarray}
\nonumber&&\frac{a^{2}}{m_{pl}^{2}}(3\alpha V+a\beta  V')\\
\nonumber&&-\dot{A}^{2}\frac{\epsilon a^{2}}{m_{pl}^{2}}(A\frac{\partial \alpha}{\partial A}+a\frac{\partial\beta}{\partial A}+\frac{3\alpha}{2})\\
\nonumber&&-\dot{a}^{2}(\frac{\epsilon A^{2}}{2m_{pl}^{2}}\alpha-\alpha+\frac{\epsilon aA^{2}}{m_{pl}^{2}}\frac{\partial \alpha}{\partial a}-2a\frac{\partial \alpha}{\partial a}
+\frac{\epsilon a A}{m_{pl}^{2}}\beta+\frac{\epsilon a^{2}A}{m_{pl}^{2}}\frac{\partial\beta}{\partial a})\\
&&-a\dot{a}\dot{A}(\frac{2\epsilon A\alpha}{m_{pl}^{2}}+\frac{\epsilon a\beta}{m_{pl}^{2}}-2\frac{\partial \alpha}{\partial A}+\frac{\epsilon A^{2}}{m_{pl}^{2}}
\frac{\partial \alpha}{\partial A}+\frac{\epsilon a^{2}}{m_{pl}^{2}}\frac{\partial\beta}{\partial a}+\frac{\epsilon a A}{m_{pl}^{2}}\frac{\partial\beta}{\partial A})=0.
\end{eqnarray}
It gives a quadratic polynomial in terms of $\dot{a}$
and $\dot{A}$, whose coefficients are partial derivatives of $\alpha$ and $\beta$ with respect to the configuration variables $a$ and $A$.
Thus the resulting expression is  equal to zero if and only if these coefficients are zero
\begin{eqnarray}
\label{1}
&&3\alpha V+a \beta V'=0,\\
\label{2}
&&2A\frac{\partial \alpha}{\partial A}+2a\frac{\partial\beta}{\partial A}+3\alpha=0,\\
\label{3}
&&\frac{\epsilon A^{2}}{2m_{pl}^{2}}\alpha-\alpha+\frac{\epsilon aA^{2}}{m_{pl}^{2}}\frac{\partial \alpha}{\partial a}-2a\frac{\partial \alpha}{\partial a}
+\frac{\epsilon a A}{m_{pl}^{2}}\beta+\frac{\epsilon a^{2}A}{m_{pl}^{2}}\frac{\partial\beta}{\partial a}=0,\\
&&\label{4}
\frac{2\epsilon A\alpha}{m_{pl}^{2}}+\frac{\epsilon a\beta}{m_{pl}^{2}}-2\frac{\partial \alpha}{\partial A}+\frac{\epsilon A^{2}}{m_{pl}^{2}}\frac{\partial \alpha}{\partial A}
+\frac{\epsilon a^{2}}{m_{pl}^{2}}\frac{\partial\beta}{\partial a}+\frac{\epsilon a A}{m_{pl}^{2}}\frac{\partial\beta}{\partial A}=0.
\end{eqnarray}

The last two equations can be simplified  by using Eq.(\ref{2}),
which are reduced to
\begin{eqnarray}
\label{b}
&&\frac{a\beta}{m_{pl}^{2}}+\frac{a^{2}}{m_{pl}^{2}}\frac{\partial \beta}{\partial a}=2\frac{\partial \alpha}{\partial A}-\frac{ \alpha A}{2m_{pl}^{2}},\\
\label{a}
&&-\alpha+\frac{aA^{2}}{m_{pl}^{2}}\frac{\partial \alpha}{\partial a}-2a\frac{\partial \alpha}{\partial a}+2A\frac{\partial \alpha}{\partial A}=0.
\end{eqnarray}
In the present paper, we will use Eqs.(\ref{1}), (\ref{2}),
(\ref{b}), (\ref{a}) as the Noether conditions with implicit
symmetries. Particularly speaking, Eq.(\ref{a})  is a partial
differential equation for $\alpha$.  We can find two solutions for
this equation, which are  $\alpha=0$ and $\alpha=C_{2}A^{1/2}$
\cite{math}. In the following subsections, combined with other
Noether conditions, we will discuss the two corresponding full
solutions in detail. Before that, we discuss the constants of motion
firstly.

Following Ref.\cite{deRitis:1990ba},  the Noether  conditions select
constants of motion. The existence of a Noether symmetry in
the model reduces the dynamics through  cyclic variables. Firstly,
we have to define the  conjugate momenta
\begin{eqnarray}
&&p_{q_{i}}=\frac{\partial\mathcal{L}_{1}}{\partial \dot{q_{i}}},
\end{eqnarray}
whose explicit expressions with different variables are
\begin{eqnarray}
&&p_{a}=\frac{\partial\mathcal{L}_{1}}{\partial \dot{a}}=3(2a\dot{a}-\epsilon\frac{aA^{2}\dot{a}+a^{2}A\dot{A}}{m_{pl}^{2}}),\\
&&p_{A}=\frac{\partial \mathcal{L}_{1}}{\partial
\dot{A}}=-\frac{3\epsilon a^{3}}{m_{pl}^{2}}(\dot{A}+ HA).
\end{eqnarray}
The equations of motion indicate $\partial\mathcal{L}_{1}/\partial
q_{i}=dp_{q_i}/dt$,  and combined with Eq.(\ref{L1}), which gives
out
\begin{eqnarray}
L_{X}\mathcal{L}_{1}=\frac{d}{dt}(\alpha p_{a}+\beta p_{A})=0.
\end{eqnarray}
Then, the required  Noether constant of motion is deduced
\begin{eqnarray}
\label{Q}
\alpha p_{a}+\beta p_{A}=Q=\mu_{0},
\end{eqnarray}
where $Q$ is the  conserved charge  with unclear physical meaning, and $\mu_{0}$ is the corresponding constant.
 In other words, a symmetry exists if at least one of
the functions $\alpha$ or $\beta$ is different from zero. As
byproducts, the constant of motion will be given out and the form of
$V(A^{2})$ is determined in correspondence to such a symmetry.

\subsection{Solution One: The Constant  Potential}\label{sub1}
The simplest solution to Eq. (\ref{a}) is $\alpha=0$. And from the
other Noether symmetry conditions, the complete solution  is
obtained \be \label{s1} \alpha=0,\,\, \beta=C_{1}a^{-1},\,\,
V=V_{0}, \ee where $C_{1}$ and $V_{0}$ are constants of integration.
This solution is equivalent to massless vector field plus a
cosmological constant, and the solution is gauge invariant. Such a
solution exists in the scalar field case as well. However, the
difference between scalar field and vector field  is that the value
of $\beta$ is a constant in the scalar field scenario, while it
varies as $a^{-1}$ in the vector field scenario. The solution
indicates that the cyclic coordinate is $a$, and  constant of motion
can be expressed as
\begin{eqnarray}
\frac{-3C_{1}a^{-1}}{m_{pl}^{2}}( a^{3}\dot{A}+ a^{2}\dot{a}A)=Q=\mu_{0},
\end{eqnarray}
which can be simplified as
\be \label{Q1} (a
A)\dot{}=a(\dot{A}+HA)=\frac{\tilde{\mu}_{0}}{a}=\frac{\mu_{0}m_{pl}^{2}}{-3C_{1}a}.
\ee

When $\tilde{\mu}_{0}=0$, Eq. (\ref{Q1}) tells us that the kinetic energy of the vector
field is zero, and the scalar field $A(t)$
evolves as $A\propto a^{-1}$. So the solution with
$\tilde{\mu}_{0}=0$  constrains the model  just a  cosmological
constant type model.

 When $\tilde{\mu}_{0}\neq0$, according to Eq.(\ref{Q1}), we get  $(\dot{A}+HA)^{2}\propto1/a^{4}$.
 The kinetic energy of the vector field behaves as a radiation field, so the solution  constrains the model
 to radiation plus cosmological constant. For a viable dark energy or inflation model,
 the potential energy must dominate, and the model again behaves like the cosmological constant model.

Since $V=V_0$, $V'=0$, the EoS parameter is
\begin{eqnarray}
\omega_{A}=\frac{\epsilon(\dot{A}+HA)^{2}/2-3V(A^{2})}{3\epsilon(\dot{A}+HA)^{2}/2+3V(A^{2})}.
\end{eqnarray}
If $\tilde{\mu}_{0}\neq0$, in the quintessence case, $\epsilon=1$, $\omega_{A} \geq -1$,
and in the phantom case, $\epsilon=-1$, $\omega_{A} \leq -1$. In both cases,
EoS parameter $\omega_{A}$ is close to $-1$ and it does not cross the phantom divider
$\omega_{A}=-1$. That is the reason why we need to consider quintessence, phantom
and quintom cases separately when the Noether symmetry is applied.

\subsection{Solution Two: The Point-Like Solution}\label{sub2}
There is  another solution which  satisfies the Noether
conditions:
\be \label{s2}
A^{2}=2\epsilon
m_{pl}^{2},\,\,\alpha=C_{2}A^{1/2},\,\,\beta=\frac{-4C_{2}A^{3/2}}{3a},\,\,V=V_{0}(\frac{A^{2}}{m_{pl}^{2}})^{9/8},
 \ee
 where $C_{2}$ and $V_{0}$ are constants. For the phantom case
where $\epsilon=-1$, $A^{2}=-2m_{pl}^{2}$, the value of the fields
are not physical, so the solution only applies to the quintessence
case. In other words, Noether symmetry may be used to select models.
The value of the vector fields are constants, and that is why we
call it the point-like solution. In this solution, the potential has
a power-law form.

Based on Eq.(\ref{s2}), the Noether constant of motion is
 \be
\label{Q2}
3C_{2}A^{1/2}(2a\dot{a}-\frac{aA^{2}\dot{a}+a^{2}A\dot{A}}{m_{pl}^{2}})+\frac{4C_{2}A^{3/2}}{am_{pl}^{2}}(a^{3}\dot{A}+
a^{2}A\dot{a})=Q=\mu_{0}.
 \ee
 When  $A^{2}=2m_{pl}^{2}$ and the value of vector field keeps as constant $\dot{A}=0$, the above
equation can be simplified as
 \be
2A^{1/2}H=\frac{\tilde{\mu}_{0}}{a^{2}},
 \ee
where $\tilde{\mu}_{0}=\mu_{0}/3C_{2}$.  The solution is $a(t)\propto t^{1/2}$, which describes the radiation era. Therefore, this solution is unable to explain the current accelerating expansion.

In summary, for the quintessence and phantom cases, the "cosmic triad" vector field scenario with
Noether symmetry behaves effectively like the cosmological constant model.


\section{Noether Symmetry Approach in Quintom Case}\label{sec4}
In this section, we are going to apply the Noether symmetry approach to the quintom case.
We look for the solution which crosses the phantom divider $\omega_{A}=-1$.
Based on the action (\ref{action2}), we can get the point-like Lagrangian
\begin{eqnarray}
\nonumber
 \mathcal{L}_{A2}&=&3a\dot{a}^{2}-\frac{3}{m_{pl}^{2}}\left(\frac{a^{3}\dot{A_{q}}^{2}+ a A_{q}^{2}\dot{a}^{2}+2 a^{2}A_{q}\dot{a}\dot{A_{q}}}{2}-a^{3}V_{q}^{2}\right)\\
 &&+\frac{3}{m_{pl}^{2}}\left(\frac{a^{3}\dot{A_{p}}^{2}+ a A_{p}^{2}\dot{a}^{2}+2 a^{2}A_{p}\dot{a}\dot{A_{p}}}{2}+a^{3}V_{p}\right).
\end{eqnarray}
The energy density and pressure for the vector
field are
\begin{eqnarray}
 &&\tilde{\rho_{A}}=\frac{3}{2}(\dot{A_{q}}+HA_{q})^{2}+3V_{q}-\frac{3}{2}(\dot{A_{p}}+HA_{p})^{2}+3V_{p},\\
 &&\tilde{P_{A}}=\frac{1}{2}(\dot{A_{q}}+HA_{q})^{2}-3V_{q}+\frac{d V_{q}}{d A_{q}}A_{q}-\frac{1}{2}(\dot{A_{p}}+HA_{p})^{2}-3V_{p}+\frac{d V_{p}}{d A_{p}}A_{p},
 \end{eqnarray}
 and the equations of motion for $A_q$ and $A_p$ are
 \begin{eqnarray}
 &&\ddot{A_{q}}+3H\dot{A_{q}}+(2H^{2}+\dot{H})A_{q}+ V'_{q}=0,\\
 &&\ddot{A_{p}}+3H\dot{A_{p}}+(2H^{2}+\dot{H})A_{p}- V'_{p}=0.
  \end{eqnarray}

The configuration space is $ \mathcal{Q}=(a,A_{q},A_{p})$, and the
generator of the Noether symmetry is
\begin{eqnarray}
\label{X}
\tilde{X}=\tilde{\alpha}\frac{\partial}{\partial a}+\tilde{\beta}\frac{\partial}{\partial A_{q}}+\gamma\frac{\partial}{\partial A_{p}}
+\dot{\tilde{\alpha}}\frac{\partial}{\partial \dot{a}}+\dot{\tilde{\beta}}\frac{\partial}{\partial \dot{A_{q}}}+\dot{\gamma}\frac{\partial}
{\partial \dot{A_{p}}},
\end{eqnarray}
where $\tilde{\alpha}$, $\tilde{\beta}$ and $\gamma$ are generic  functions of $a$, $A_{p}$, $A_{q}$.
 Noether symmetry requires $L_{\tilde{X}} (\mathcal{L}_{A2}+ \mathcal{L}_{m})=0$, so
 the Noether conditions are
\begin{eqnarray}
\label{q1}
&&3\tilde{\alpha}( V+\tilde{V'})+a \tilde{\beta} V_{q}'+a \gamma V_{p}'=0,\\
\label{q2}
&&2A_{q}\frac{\partial \tilde{\alpha}}{\partial A_{q}}+2a\frac{\partial\tilde{\beta}}{\partial A_{q}}+3\tilde{\alpha}=0,\\
\label{q21}
&&2A_{p}\frac{\partial \tilde{\alpha}}{\partial A_{p}}+2a\frac{\partial\gamma}{\partial A_{p}}+3\tilde{\alpha}=0,\\
&&\label{q3}
\frac{2 A_{q}\tilde{\alpha}}{m_{pl}^{2}}+\frac{ a\tilde{\beta}}{m_{pl}^{2}}-2
\frac{\partial \tilde{\alpha}}{\partial A_{q}}+\frac{ A_{q}^{2}}{m_{pl}^{2}}\frac{\partial \tilde{\alpha}}
{\partial A_{q}}+\frac{ a^{2}}{m_{pl}^{2}}\frac{\partial\tilde{\beta}}{\partial a}+\frac{ a A_{q}}{m_{pl}^{2}}\frac{\partial\tilde{\beta}}{\partial A_{q}}=0,\\
&&\label{q31}
\frac{2 A_{p}\tilde{\alpha}}{m_{pl}^{2}}+\frac{ a\gamma}{m_{pl}^{2}}+2\frac{\partial \tilde{\alpha}}{\partial A}+
\frac{A_{p}^{2}}{m_{pl}^{2}}\frac{\partial \tilde{\alpha}}{\partial A_{p}}+\frac{ a^{2}}{m_{pl}^{2}}\frac{\partial\gamma}
{\partial a}+\frac{a A_{p}}{m_{pl}^{2}}\frac{\partial\gamma}{\partial A_{p}}=0,\\
\label{q4}
\nonumber
&&\frac{ A_{q}^{2}}{2m_{pl}^{2}}\tilde{\alpha}-\tilde{\alpha}+\frac{ aA_{q}^{2}}{m_{pl}^{2}}\frac{\partial \tilde{\alpha}}
{\partial a}-2a\frac{\partial \tilde{\alpha}}{\partial a}
+\frac{ a A_{q}}{m_{pl}^{2}}\tilde{\beta}+\frac{ a^{2}A_{q}}{m_{pl}^{2}}\frac{\partial\tilde{\beta}}{\partial a}\\
&&-\frac{ A_{p}^{2}}{2m_{pl}^{2}}\tilde{\alpha}-\frac{ aA_{p}^{2}}{m_{pl}^{2}}\frac{\partial \tilde{\alpha}}{\partial a}-
\frac{ a A_{p}}{m_{pl}^{2}}\gamma-\frac{ a^{2}A_{p}}{m_{pl}^{2}}\frac{\partial\gamma}{\partial a}=0.
\end{eqnarray}
The above equations have an obvious solution which is
\begin{eqnarray}
\label{s3}
&&\tilde{\alpha}=0,\,\,\tilde{\beta}=Ca^{-1},\,\,\gamma=\tilde{C}a^{-1},\\
\label{s3v} &&\tilde{C}V_{p}'=-CV_{q}',
\end{eqnarray}
where $C$ and $\tilde{C}$ are both constants of integration, and at
least one of the three parameters $\tilde{\alpha}$, $\tilde{\beta}$,
$\gamma$ is not zero. In the following, according to the values of
$C$, $\tilde{C}$ and $\mu_{0}$, we will discuss the various
solutions .

\subsection{$C=0$ and $\tilde{C}\neq0$}
In the case of  $C=0$ and $\tilde{C}\neq0$,  we get $V_{p}'=0$,
and the conserved charge is
\begin{eqnarray}
\label{Q2} \nonumber &&Q= \tilde{\alpha} p_{a}+\tilde{\beta}
p_{A_{q}}+\gamma p_{A_{p}}
=\tilde{C}\frac{3a^{2}}{m_{pl}^{2}}(\dot{A_{p}}+A_{p}H) =\mu_{0},
\end{eqnarray}
and both $a$ and $A_{q}$ are cyclic variables. When $\mu_{0}\neq0$,
the above equations tell us that the phantom type field is the same
as that from the constant potential solution, while the quintessence
type field is free of the constraint. This solution allows that one
crosses  the phantom divider $\omega_{A}=-1$. When  $\mu_{0}=0$, the
solution constrains the model to just a quintessence field plus the
cosmological constant. This solution does not allow that one crosses
the phantom divider $\omega_{A}=-1$. The case $\tilde{C}=0$ and
$C\neq0$ can be treated exactly in the same way. And the results are
similar, except for that the role of quintessence field is replaced
by the phantom type field.

\subsection{$C\neq0$ and $\tilde{C}\neq0$}
If both the parameters $C$ and $\tilde{C}$ are not zero, Eq.
(\ref{s3v}) gives a  constraint on the form of  potentials which is
related to the first derivative of  potentials.  The conserved
charge is
\begin{eqnarray}
\label{Q2}
-C\frac{3a^{2}}{m_{pl}^{2}}(\dot{A_{q}}+A_{q}H)+\tilde{C}\frac{3a^{2}}{m_{pl}^{2}}(\dot{A_{p}}+A_{p}H)=Q=\mu_{0}.
\end{eqnarray}

 Here, when $\mu_{0}\neq0$, we discuss a particular situation that is  both $\dot{A_{q}}+HA_{q}$ and $\dot{A_{p}}+HA_{p}$ are proportional to $a^{-2}$,
 which indicates that the kinetic energy of both fields behaves like a radiation field.
Substituting the results into the equations of motion, we find that
$V'_{q}=V'_{p}=0$, which means the behaviors of $A_{p}$ and $A_{q}$
are the same as those discussed in Section \ref{sub1}.

When  $\mu_{0}=0$, Eq. (\ref{Q2}) tells us that
\begin{eqnarray}
\label{pq}
(\dot{A_{p}}+A_{p}H)=\frac{C}{\tilde{C}}(\dot{A_{q}}+A_{q}H).
\end{eqnarray}
 Combining   Eqs. (\ref{s3v}) and (\ref{pq}), we get
  \begin{eqnarray}
\tilde{\omega}_{A}=\frac{\tilde{P}_{A}}{\tilde{\rho}_{A}}=\frac{\frac{1}{2}(1-\frac{C^{2}}{\tilde{C}^{2}})(\dot{A_{q}}+HA_{q})^{2}+(1-\frac{C}{\tilde{C}})\frac{d
V_{q}}{d
A_{q}}A_{q}-3V_{q}-3V_{p}}{\frac{3}{2}(1-\frac{C^{2}}{\tilde{C}^{2}})(\dot{A_{q}}+HA_{q})^{2}+3V_{q}+3V_{p}}.
 \end{eqnarray}
Obviously, when $C=\tilde{C}$, $\tilde{\omega}_{A}=-1$. If we
require $\tilde{\omega}_{A}<(>)-1$, the following condition must be
satisfied
 \be
0<(>)(\frac{\tilde{C}}{C}-1)\left(2(\frac{\tilde{C}}{C}
+1)(\dot{A_{q}}+HA_{q})^{2}+V_{q}'A_{q}\right).
 \ee
Therefore this model allows that one crosses the phantom divider
$\tilde{\omega}_{A}=-1$ as expected. We take the following two
examples to illustrate this point
\begin{eqnarray}
&&Example\,\,One:\,\,\,V_{q1}=V_{0}-\frac{1}{2}m_{1}^{2}A_{q}^{2},\,\,V_{p1}=V_{0}+\frac{1}{2}\tilde{m}_{1}^{2}A^{2},\,\,\frac{m_{1}^{2}}{\tilde{m}^{2}_{1}}=\frac{\tilde{C}}{C};\\
&&Example\,\,Two:\,\,\,V_{q2}=V_{0}+\frac{1}{2}m_{2}^{2}A^{2},\,\,V_{p1}=V_{0}+\frac{1}{2}\tilde{m}_{2}^{2}A^{2},\,\,\frac{m_{2}^{2}}{\tilde{m}^{2}_{2}}=-\frac{\tilde{C}}{C}.
\end{eqnarray}
 In  the first example, without loss of generality, assuming $\tilde{C}/C>1$, when  $2(\tilde{C}/C
+1)(\dot{A_{q}}+HA_{q})^{2}+V_{q}'A_{q}$ crosses zero,
$\tilde{\omega}_{A}$ crosses over $\tilde{\omega}_{A}=-1$. It is not
surprising  that the quintessence-like field takes the dominant
role. In the second example, if
 $\tilde{C}/C\ll-1$, or $m_{2}^{2}\ll \tilde{m}_{2}^{2}$,  it is easy to get
 $\tilde{\omega}_{A}<-1$.

The most interesting point is that the solution does not require a
particular form for the potentials, so we have the freedom of
choosing  potentials by observations. Furthermore, this kind of
constraint is coming from symmetries of the system.


\section{Conclusion}\label{sec5}

In this paper, we have studied the choice of potentials in ``cosmic
triad" vector field  model by using Noether symmetry approach. The
existence of Noether symmetry implies that the Lie derivative of the
Lagrangian with respect to the related infinitesimal generator
vanishes.  The phase space is  constructed by taking the scale
factor $a$ and the field $A$ as independent dynamical variables. In
the configuration space which is spanned by $a$ and $A$, the
point-like Lagrangian of the model  is constructed such that its
variations with respect to these dynamical variables yield correct
field equations. Then, the dynamical system is simplified with
Noether symmetry.

The Noether symmetry is used to select a class of potentials.
 We have derived the Noether conditions for three different vector field models.
In the quintessence and phantom cases,  solutions with constant
potential have been obtained. And, a point-like solution with
power-law potential exists for quintessence case only. This suggests
that we may use Noether symmetry to select theoretical models. In
the quintom case, we find that the Noether symmetry requires
$\tilde{C}V_{p}'=-CV_{q}'$. This result gives a useful constraint on
the quintom potentials, and the solution has the desired crossing
over $\omega_{A}=-1$ behavior.

\section*{Acknowledgements}
We thanks Rong-gen Cai, Hui Li,  Ya Tian and Nan Liang for useful
discussions. We also thanks the anonymous referees for their useful
suggestions. This work was supported by CQUPT under Grant No.
A2009-16, the National Basic Research Program of China (973 Program)
under grant Nos. 2007CB815401 and 2010CB833004, the National Natural
Science Foundation of China key project under grant Nos. 10533010
and 10935013,  the Distinguished Young Scholar Grant 10825313, and
the Natural Science Foundation Project of CQ CSTC under grant No.
2009BA4050.

\end{document}